\documentclass[twocolumn,traditabstract]{aa}
\usepackage{txfonts}
\usepackage{graphicx}
\usepackage{natbib}
\usepackage{hyperref}
\usepackage[running,switch]{lineno}

\newcommand{\kms}{\mbox{km~s$^{-1}$}}
\newcommand{\mols}{\mbox{molec.~s$^{-1}$}}

\begin{document}

\title{Perihelion observations of interstellar comet 3I/ATLAS with the IRAM 30-m
  telescope
\thanks{Based on observations carried out with the IRAM 30-m telescope.
IRAM is supported by INSU/CNRS (France), MPG (Germany) and IGN (Spain).}
}
 
\author{N. Biver\inst{1}
   \and D. Bockel\'ee-Morvan\inst{1}
   \and R. Moreno\inst{1}
   \and J. Crovisier\inst{1}
   \and G. Paubert\inst{2}
   \and V. Zakharov\inst{1}
   \and J. Boissier\inst{3}
   \and M.~A. Cordiner\inst{4,5}
   \and N.~X. Roth\inst{4,6}
   }
\institute{LIRA, Observatoire de Paris, PSL Research University, CNRS, 
      Sorbonne Universit\'e, Universit\'e de Paris, 
      5 place Jules Janssen, F-92195 Meudon, France
 \and IRAM, Avd. Divina Pastora, 7, 18012 Granada, Spain
 \and IRAM, 300, rue de la Piscine, F-38406 Saint Martin d'H\`eres, France
 \and NASA Goddard Space Flight Center, 8800 Greenbelt Road, Greenbelt, MD 20771, USA
 \and Department of Physics, Catholic University of America, 620 Michigan Ave. NE, Washington, DC 20064, USA
 \and Department of Physics, American University, 4400 Massachusetts Ave NW, Washington, DC 20016, USA
} 
   \titlerunning{Observations of comet 2I with IRAM}
   \authorrunning{Biver et al.}
   \date{\today}

   \abstract{3I/ATLAS is the third interstellar comet identified as passing through the Solar System. Its high outgassing activity and favourable perihelion passage on October 29, 2025 UT provided an excellent opportunity to investigate the composition of its coma gases through millimeter spectroscopy.  We present observations undertaken with the IRAM 30-m telescope on November 1--3, 2025 at an heliocentric distance of 1.36--1.37 au. Lines of HCN, CH$_3$OH, CO, and H$_2$CO are well detected, and $\sim$4$\sigma$ detections are obtained for CS and CH$_3$CN. The search for H$_2$S was unsuccessful. Abundances of CO, H$_2$CO, CH$_3$OH, and CH$_3$CN relative to HCN are in the upper ranges of values measured in Solar System comets. The sulfur-to-carbon abundance ratio in 3I/ATLAS's coma is at most the minimum value observed in comets.  The unusually low expansion velocity of coma gases suggests a near-nucleus gas flow driven by heavy molecules such as CO$_2$, and/or a large fraction of the gaseous production coming from subliming icy grains.      }


\maketitle

\section{Introduction}
Comets are fingerprints of the formation of planetary systems.   Water and the numerous complex molecules composing the ices were likely synthesized in the molecular cloud precursor of the Solar System \citep{Altwegg2019,2022arXiv220613270C,Biv24}. What are the physical and chemical properties of icy planetesimals in other planetary systems and how do they compare with Solar System primitive bodies? For such investigation, the passing of a bright cometary-like interstellar object (ISOs) through the Solar System was long awaited. The first interstellar object, 1I/'Oumuamua, was discovered in October 2017 a few days after its closest approached to Earth. 
 There was no detection of coma gases and dust particles \citep[][and references therein]{2019NatAs...3..594O}, but astrometric measurements suggested the presence of non-gravitational forces related to cometary activity \citep{Micheli2018}. The second interstellar object discovered in 2019 (2I/Borisov) showed cometary activity and was extensively observed. Several simple volatile species commonly observed in Solar System comets (OH, C$_2$, CN, HCN, NH, NH$_2$, CO, Ni, Fe) were detected \citep[e.g.][]{2019A&A...631L...8O,2021A&A...650L..19O,2020NatAs...4..861C,2025arXiv250705051D}. Spectroscopic properties and mixing ratios were found remarkably similar to Solar System comets. 2I/Borisov was found to be carbon-chain depleted, similar to the class of
carbon-chain depleted comets of Solar System origin \citep{2019A&A...631L...8O,2020ApJ...889L..38K,2020ApJ...889L..30L}. However, a high enrichment in CO relative to H$_2$O was revealed, suggesting specific formation conditions in its natal protoplanetary disk \citep{2020NatAs...4..861C,2020NatAs...4..867B}.



In this letter, we report on spectroscopic observations with the 30-m telescope of the Institut de radioastronomie millim\'etrique (IRAM) of the third identified ISO, 3I/ATLAS (also named 3I/2025 N1 (ATLAS)) which were performed  near its perihelion on October 29.48, 2025 UT, at $r_h$ = 1.356 au from Sun. This interstellar object was discovered with the robotic telescope ATLAS in Chile on July 1, 2025 as it was at 5 au from the Sun, and confirmed to
be a comet the day after \citep[CBET 5578,][]{2025ApJ...989L..36S}. Subsequent studies have revealed a fast increase of its gaseous activity as it approached the Sun, making possible the detection of molecular species at millimeter wavelengths. Detections of HCN and CH$_3$OH were reported from observations with the James Clerk Maxwell Telescope (JCMT) at $r_h$ = 2.1 au \citep{2025arXiv251002817C} and $r_h$ = 1.39 au \citep{2025CBET.5628....1K}, respectively. HCN and CH$_3$OH molecular lines were mapped with the ALMA Compact Array (ACA) in a time range spanning 2.6--1.7 au pre-perihelion from which it was inferred that the CH$_3$OH/HCN ratio in 3I/ATLAS is among the largest values measured in any comet \citep{2025arXiv251120845R}. In this Letter, we present the detection of HCN, CH$_3$OH, CO, H$_2$CO, CH$_3$CN and CS at $r_h$ = 1.36--1.37 au (November 1--3, 2025). Sensitive upper limits for several molecules, including HNC, H$_2$S, are also obtained. We then compare abundance ratios in 3I/ATLAS to values measured in Solar System comets and discuss the unique dynamic properties of the coma gases.

\section{Observations}

The observations were conducted with the IRAM 30-m telescope (Sierra Nevada, Spain) from November 1 to 3, 2025 as Director Discretionary Time (DDT D05-25). Comet 3I/ATLAS was at that time at r$_h$ $\sim$ 1.36 au from the Sun, and $\Delta$ $\sim$ 2.27 au from Earth (Table~\ref{tablog}).
The EMIR 150 GHz and 230 GHz dual-side band receivers were used, with two distinct tunings for the latter. Each of the three setups covered 8 GHz in each side band.
The setup with the 150 GHz receiver targetted the H$_2$S 1$_{\rm 10}$-- 1$_{\rm 01}$ (168.8 GHz) and CS $J$(3--2) (147 GHz) lines. One 230 GHz setup was optimized to look for HCN $J$(3--2) at 265.9 GHz in the upper side band (USB) and lines of the CH$_3$OH 252 GHz multiplet in the lower sideband (LSB). The other 230 GHz setup was designed to target CO $J$(2--1) (230.5 GHz) and H$_2$CO 3$_{\rm 12}$--2$_{\rm 11}$ (225.7 GHz) in LSB,  and some of the strongest methanol lines near 242 GHz and CS $J$(5--4) (244.9 GHz) line in USB. As backends, we used the Fast Fourier Transform Spectrometer (FTS, resolution of 200 kHz) and VESPA autocorrelator (resolution of 40 kHz) in parallel.  

The three setups were executed each day (Table~\ref{tablog}). The first day offered low opacity conditions (below 2~mm precipitable water vapour (pwv)) but the weather was unstable and windy (resulting to an 1.1~h-long interruption). The next day started with a high opacity (11mm pwv) which decreased with time, but a good focus and pointing was again difficult to obtain,
especially at the beginning of the observations during sunrise when the source was at low elevation. On the third day
the opacity was relatively good (3-4~mm pwv) and the pointing and focus were generally more
reliable. The pointing errors, including those related to the used ephemeris, were estimated afterwards and found to be $\sim$3\arcsec~for most observations (Table~\ref{tabarea}). The telescope beam size is 10.8\arcsec~at 230.5 GHz.

\begin{figure}[!h]
\centering
\resizebox{\hsize}{!}{\includegraphics[angle=270]{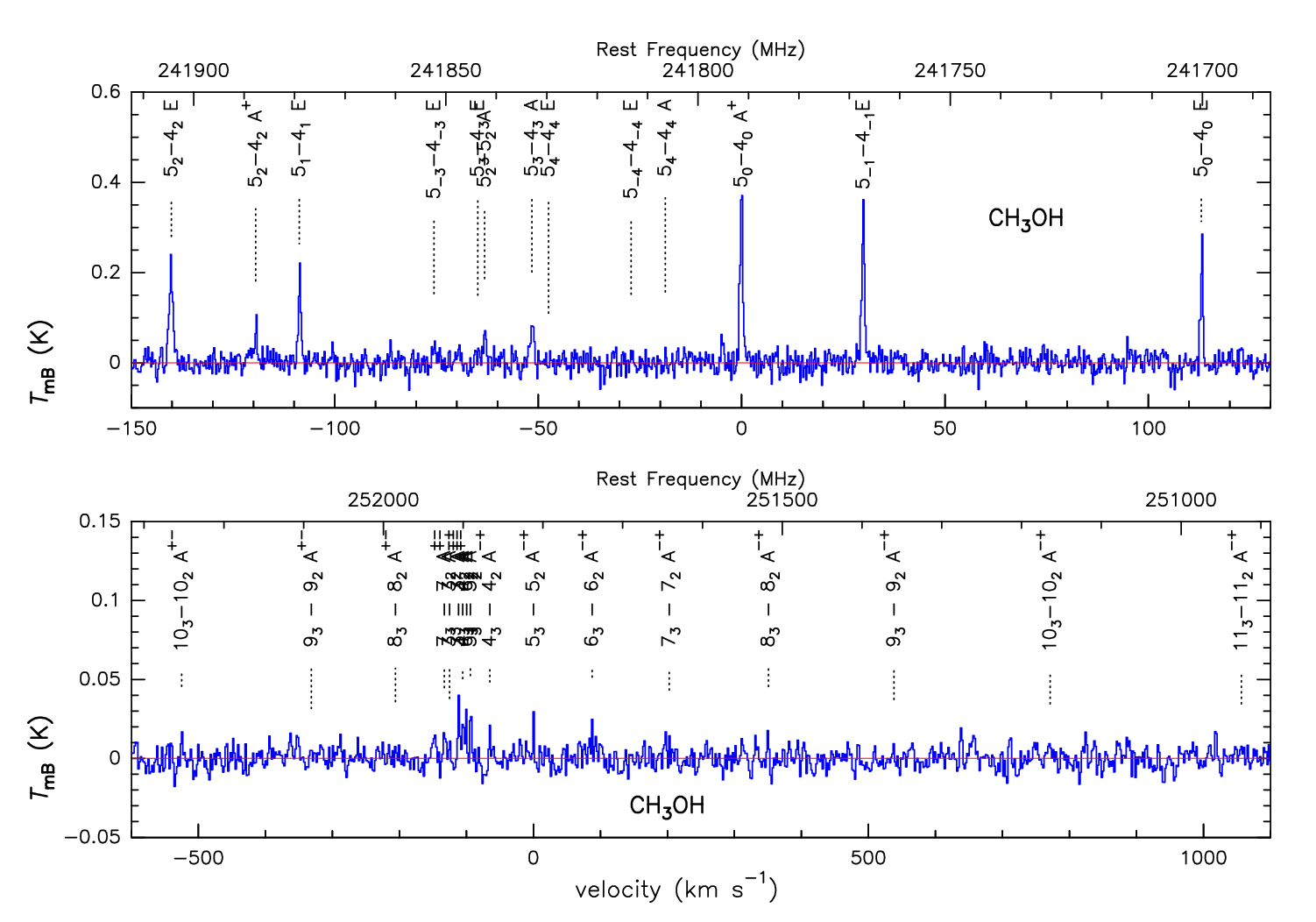}}
\caption{Methanol lines near 242 GHz and 252 GHz observed in comet 3I/ATLAS on November 1--3, 2025.\label{figmet}}
\end{figure}

\begin{figure}
\resizebox{\hsize}{!}{\includegraphics[angle=270,width=12cm]{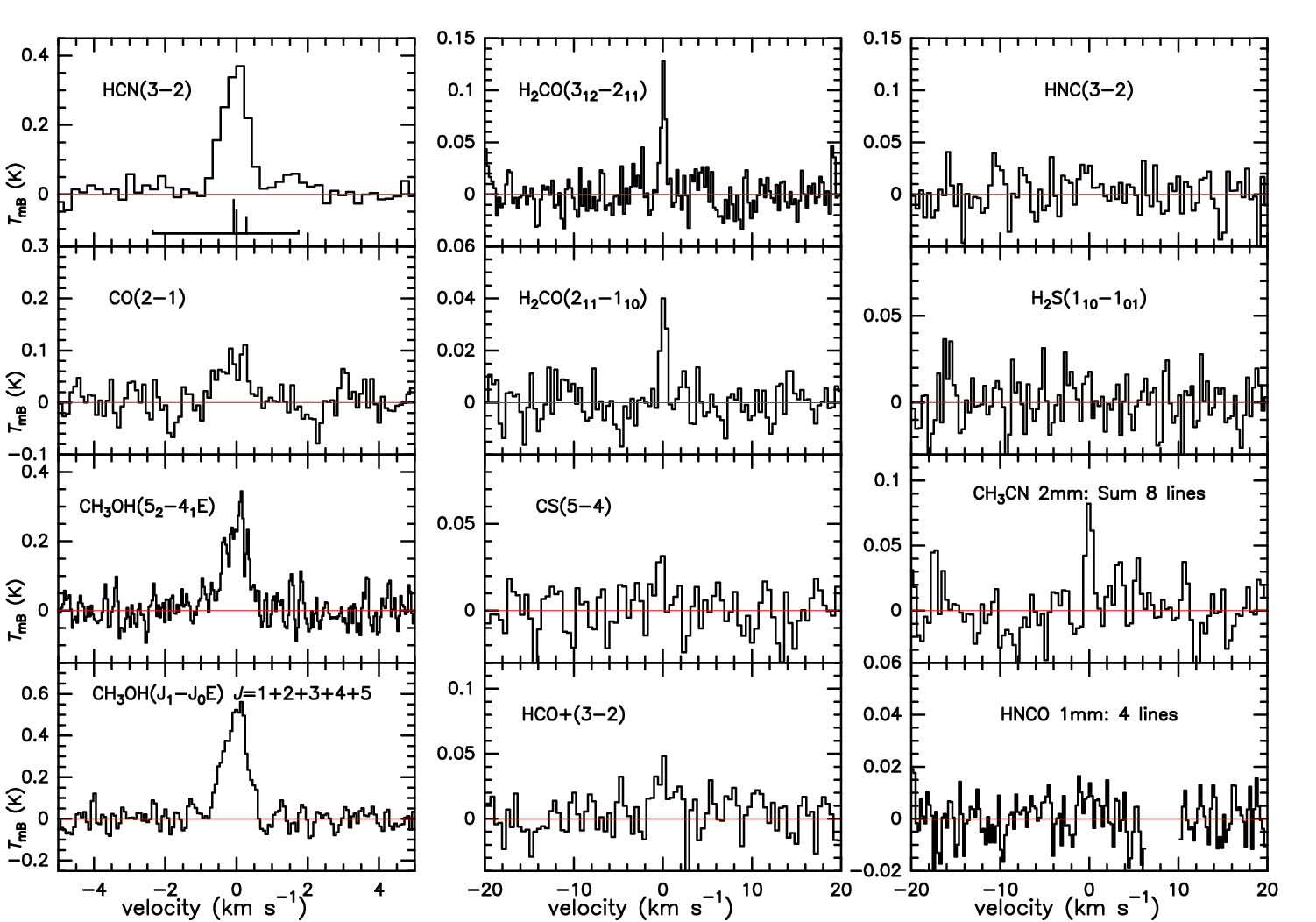}}
\caption{Sample of spectra obtained in comet 3I/ATLAS on November 1--3, 2025. Line characteristics are given in Table~\ref{tabarea}. The plot with the HCN $J$(3--2) spectrum shows the position and relative intensities of the hyperfine components. \label{figsp}} 

\end{figure}

\section{Results}
\label{results}

Figures \ref{figmet},  \ref{figsp} show sample spectra of comet 3I/ATLAS. A list of detected lines is given in Table~\ref{tabarea}, together with  their line areas in main-beam brightness temperature scale. HCN, CH$_3$OH, CO, and H$_2$CO lines are well detected. CH$_3$OH is detected through multiple lines ($>$30), including with the 150 GHz receiver. CS lines are marginally detected, at 3.3$\sigma$ for $J$(5--4), and 2.3$\sigma$ for $J$(2--1). CH$_3$CN $J$(8--7) (147 GHz) and $J$(9--8) (166 GHz) lines are detected at 3$\sigma$ and 2.5$\sigma$, respectively. Combining the different lines, CS and CH$_3$CN are detected at 4$\sigma$. A marginal detection, slightly below 3$\sigma$ is also obtained for HNC. The $J$(3--2) HCO$^+$ line (267.6 GHz) is also marginally seen. H$_2$S is not detected. The large bandwidth of EMIR receivers allowed us to search for many species, including using line stacking, so Tables~\ref{tabarea} and \ref{tababund} include upper limits for many species observed in cometary atmospheres, including deuterated species.

In order to derive production rates, we followed the approach used in previous papers \citep[e.g.][and references therein]{2021A&A...648A..49B,2024A&A...690A.271B}. The expansion
velocity was estimated from the line profiles. The HCN, CO and CH$_3$OH lines observed at high spectral resolution
have similar, slightly blueshifted ($\Delta v\approx-0.05$ \kms~on average), line shapes (Fig.~\ref{figsp}, Table~\ref{tabarea}). 
Gaussian fits to the lines yielded on average half-widths of $VHM=0.54\pm0.04$ and $0.34\pm0.02$ \kms, in the blue and red wings of the profiles, respectively (Table~\ref{tableVHM}).  A model considering asymmetric outgassing with expansion velocities of 0.42 and 0.32~\kms~in the sunward and antisunward hemispheres provided a good fit to the line profiles (Fig.~\ref{linefits}). Isotropic outgassing at an expansion velocity of 0.37 \kms yields the same production rates, so this approximation was used.

In the excitation model, which considers both collisional excitation  by water and electrons and radiative processes, we assumed a constant kinetic temperature  and an electron-density scaling factor $x_{ne}$ = 0.5 \citep{1997PhDT........51B}. With a kinetic temperature of~60 K, the rotational temperatures derived from the modeled CH$_3$OH line intensities are consistent with the values derived from the observations of the CH$_3$OH multiplets  (Fig.~\ref{figmet}): $T_{\rm rot}$(165 GHz)=$44.5\pm3.5$ K, $T_{\rm rot}$(252 GHz)=$48\pm5$ K (see rotation diagram in Fig.~\ref{figdiag}), $T_{\rm rot}$(242 GHz)=$29\pm1.5$ K. In Solar System comets, the spatial distribution of H$_2$CO cannot be explained by a direct release from the nucleus and suggests production from the dust grains. For this species, we adopted a Haser daughter distribution to describe the radial evolution of the H$_2$CO number density. The parent scale length was set to $L_p$ = 2000 km (i.e., 0.5 times the H$_2$CO photodissociative scalelength), based on measurements in Solar System comets \citep{2021ApJ...921...14R,2024A&A...690A.271B}. The derived H$_2$CO production rate is not significantly dependent on the assumed $L_p$ since the beam size is large ($\sim$ 17000 km diameter at the frequency of the H$_2$CO line) \footnote{ACA observations suggest that H$_2$CO is more extended in 3I/ATLAS (M. Cordiner, personal communication). A production rate 9\% and 18\% higher than given in Table~\ref{tababund} is obtained for $L_p$ = 6000 and 8000 km, respectively.}.  For SO and CS, we assumed $L_p$ = 3000 km and 1700 km, respectively  \citep{2024A&A...690A.271B}. For other molecules (except HCO$^+$, that we do not consider in the analysis), we assumed direct release from the nucleus. Derived production rates (or their upper limits) are listed in Table~\ref{tababund}. 


The determination of mixing ratios with respect to water relies on contemporaneous observations of H$_2$O or of its photodissociation products (OH, H).   \citet{DBMMAJIS} report $Q_{\rm H_2O}$ = (6.58$\pm$0.13)$\times10^{28}$~\mols~on November 2 from infrared H$_2$O data with a 3$\times$3\arcmin~FOV obtained with the Moons And Jupiter Imaging Spectrometer (MAJIS) onboard the Jupiter Icy Moons Explorer (JUICE). On the other hand \citet{Combi2025}, from Lyman-$\alpha$ data obtained with SOHO/SWAN, report a large water production rate $Q_{\rm H_2O} = 3.2 \times10^{29}$~\mols~at 1.4 au post-perihelion (November 6) \footnote{\citet{Tan2026} report $Q_{\rm H_2O}$ values 5--6 times smaller using the same SOHO/SWAN data set, but also conclude that 3I/ATLAS is an hyperactive comet. Still unpublished MAVEN data are consistent with results from \citet{Combi2025} (J. Deighan, personal communication).}. SOHO/SWAN's large field-of-view (FOV) was possibly sensitive to large-scale water production from subliming icy grains. Extended water production from icy grains (i.e., hyperactivity) at perihelion is confirmed from the icy area fraction \citep[see][]{Lis2019} of $\sim$900\% at 1.4 au derived from the $Q_{\rm H_2O}$ measured by SOHO/SWAN and setting a nucleus radius of 1.3 km as derived from HST observations \citep[][]{2026arXiv260121569H}. For interpreting the IRAM data ($\sim$11\arcsec~FOV), we used the $Q_{\rm H_2O}$ value of $6.6\times10^{28}$~\mols~deduced from the MAJIS data.
Observations of OH 18-cm lines at the Nan\c{c}ay radio telescope (3.5\arcmin$\times$19\arcmin~beam) yielded $Q_{\rm OH}$ = (7.0$\pm$0.6)$\times10^{28}$~\mols~ for October 13--19 ($r_h$ = 1.43 au) \citep{Cro25}\footnote{Updated value from a reevaluation of the maser inversion.}, and $Q_{\rm OH}$ = (6.1$\pm$1.4)$\times10^{28}$~\mols~for November 18--22 ($r_h$ = 1.57 au).


Derived production rates and mixing ratios can be compared to pre-perihelion measurements, when the comet was farther from the Sun. The CO/H$_2$O mixing ratio at $r_h$ = 1.36 au ($\sim$ 10\%) is much lower than the value at 3.32 au ($\sim$ 165\%) derived from JWST observations \citep{2025ApJ...991L..43C}. This is not surprising as water sublimation is poorly efficient at $r_h \geq 3$ au. The CH$_3$OH production rate at $r_h$ = 1.36 au is a factor of 2 lower than the trend with $r_h$ established from ACA observations in the range 2.6--1.7 au \citep{2025arXiv251120845R}, suggesting that 3I/ATLAS activity levelled off when approaching perihelion. The CH$_3$OH/H$_2$O ratio (5.1$\pm$0.1\%) is within 1.5$\sigma$~consistent with the pre-perihelion ($r_h$ = 1.43 au) value of 8$\pm$2\% derived by \citet{2025arXiv251120845R}.

\begin{table*}
\caption[]{Production rates and abundances in comet 3I/ATLAS on November 1--3, 2025.}\label{tababund}
\begin{center}
\begin{tabular}{lccccc}
\hline\hline
 Molecule & Prod. Rate\tablefootmark{a} & \multicolumn{2}{c}{Abundance relative to water} & \multicolumn{2}{c}{Abundance relative to HCN}\\
      &  ($10^{25}$~molec.s$^{-1}$) & 3I/ATLAS\tablefootmark{b}         & Comets  & 3I/ATLAS  & Comets\\
\hline
  HCN       &   $4.1\pm0.2$  & $0.062\pm0.003$\%  & 0.08--0.25\%    &  1    & 1      \\
  HNC       &   $0.5\pm0.2$\tablefootmark{c}  &  $<0.009$\%       & 0.0015--0.035\%\tablefootmark{d}  &  $<0.13$  &0.01--0.29\tablefootmark{d}     \\
  CH$_3$CN  &   $1.5\pm0.4$  & $0.023\pm0.006$\%  & 0.008--0.054\%  &  $0.36\pm0.09$ & 0.07--0.52\\
  CH$_3$OH  & $338\pm5$      & $5.1\pm0.1$\%      & 0.4--6.1\%      &  $83\pm4$  & 4--64\tablefootmark{e}       \\
  H$_2$CO   &  $35\pm4$\tablefootmark{c}      & $0.53\pm0.06$\%    & 0.13--1.4\%     &  $8.6\pm1.0$  & 1.3--13 \\
  CO        & $680\pm11$     & $10.3\pm1.7$\%      & 0.4--35\%       &  $167\pm8$  &8--270\tablefootmark{e}   \\
  H$_2$S    &  $<39$         &  $<0.59$\%         & 0.09--1.5\%     &  $<9$        & 1.5--9  \\
  CS        &  $1.7\pm0.4$\tablefootmark{c}   & $0.024\pm0.006$\%  & 0.05--0.20\%\tablefootmark{d}    &  $0.34\pm0.11$ & 0.2--3.2\tablefootmark{d}\\
  HC$_3$N   &  $<2.1$        &  $<0.031$\%        & 0.002--0.068\%  &  $<0.52$       & 0.016--0.42\\
  OCS       &  $<50$         &  $<0.75$\%         & 0.05--0.40\%    &  $<12$        & 0.43--3.5 \\
  CH$_3$CHO & $14\pm7$       &  $<0.32$\%         & 0.04--0.08\%    &  $<5.1$       & 0.1--0.8 \\
  (CH$_2$OH)$_2$ & $33\pm13$ & $<0.59$\%          & 0.07--0.35\%    &  $<9.5$       & 0.8--2.9 \\
  SO        &  $<27$\tablefootmark{c}         & $<0.41$\%          & $<$0.03--0.30\% &  $<7$  &0.3--4.6       \\
  HNCO      & $25\pm11$      & $<0.52$\%          & $<$0.01--0.62\% &  $<8$    & 0.2--4.8     \\
  HCOOH     & $<109$         & $<1.7$\%           & $<$0.04--0.68\% &  $<27$  & 0.6--7.2      \\
  NH$_2$CHO & $<4.8$         & $<0.08$\%          & 0.015--0.022\%  &  $<1.2$ & 0.09--0.3       \\
  HDO       & $<121$         & $<1.83$\%          & 0.028--0.12\%  &  -- & --  \\
  CH$_3$OD  & $<17.9$        & $<0.27$\%          & $<0.02$\%      & -- & --\\
\hline
\end{tabular}
\end{center}
\tablefoot{\tablefoottext{a}{ Uncertainties correspond to propagated 1$\sigma$ uncertainties on the line areas, and do not include an estimated additional uncertainty of 7\% related to model parameters.}\tablefoottext{b}{Assuming the water production rate $Q_{\rm H_2O}$ = $6.58\times10^{28}$~\mols~measured from MAJIS/JUICE \citep{DBMMAJIS}. Uncertainties in $Q_{\rm H_2O}$ are not taken into account.}
  \tablefoottext{c}{Daughter distributions with $L_p$=0--1600, 1700, 2000, and 3000~km are assumed for HNC, CS, H$_2$CO and SO, respectively. Due the large beam size (15000--26000~km), distributed sources with Lp$\leq$3000~km provide similar results.}\tablefoottext{d}{Decreasing abundance  with increasing $r_h$.} \tablefoottext{e}{Excluding values in C/2016 R2 (PanSTARRS): 280, 2.6$\times$10$^4$ for CH$_3$OH/HCN and CO/HCN, respectively.}
}\\
\end{table*}

\begin{figure}[h]
\resizebox{\hsize}{!}{\includegraphics[angle=270,width=12cm]{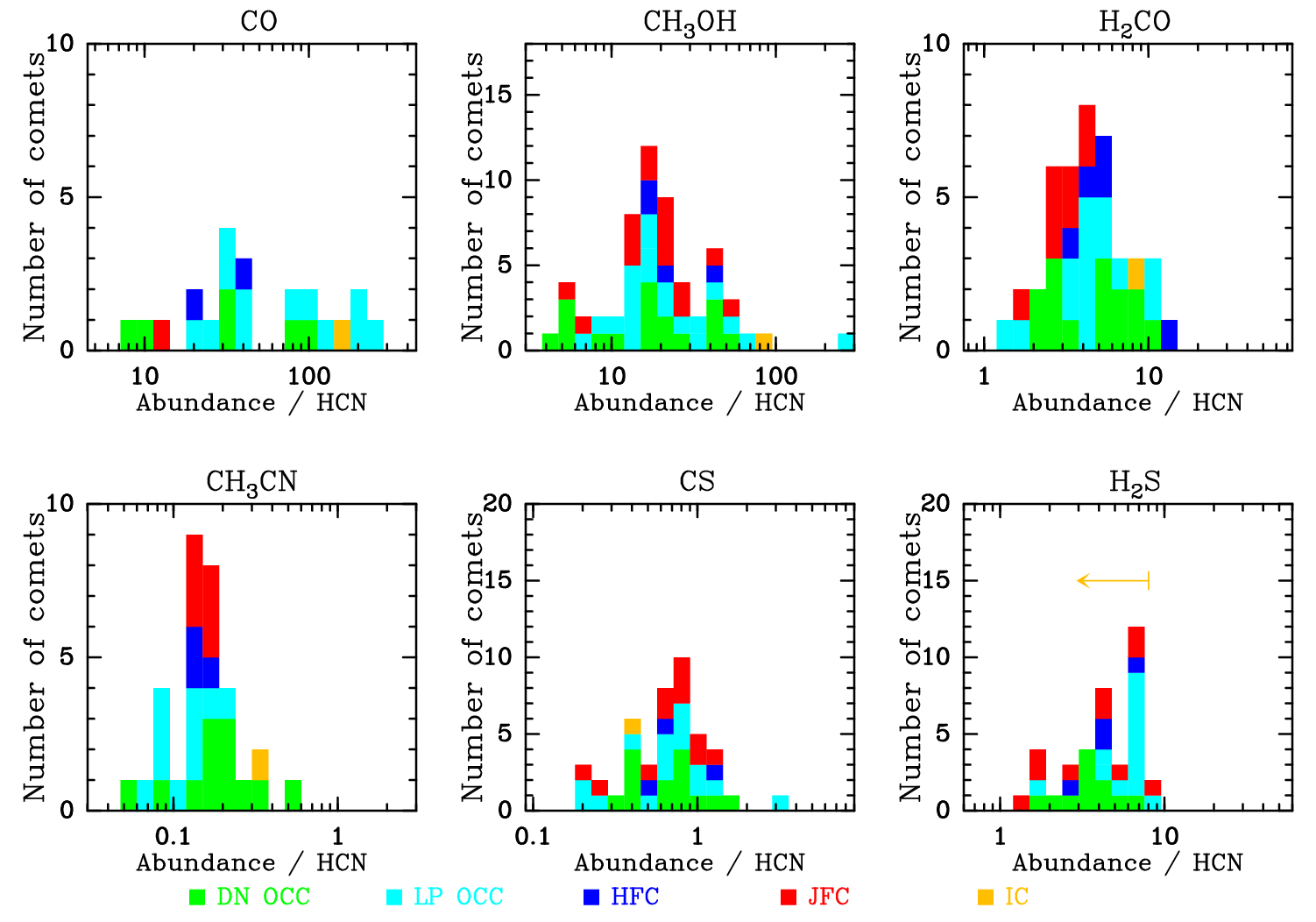}}
\caption{Histograms of the abundances relative to HCN measured from observations at radio wavelengths. Dynamically new (DN) and long-period (LP) comets from the Oort cloud (OCC), Halley-family (HFC) and Jupiter-family (JFC) comets, are indicated in green, cyan, blue, and red colors, respectively. The orange color is for 3I/ATLAS. See \citet{Biv24} and references therein.  } 
\label{figabunHCN}
\end{figure}



\section{Discussion}
The molecular species detected in this millimeter spectral survey of 3I/ATLAS are those which are commonly observed in Solar System comets \citep[e.g.][]{2024A&A...690A.271B,Biv24}, indicating that the parent protoplanetary disk (and natal molecular cloud) of this icy object share the same chemical properties, at least in first approximation. However, the derived abundances in 3I/ATLAS's coma are somewhat different from those of Solar System comets (Table~\ref{tababund}). The first striking difference is the low ($\sim$ 0.06\%) HCN/H$_2$O mixing ratio. For comparison, the HCN/H$_2$O ratio measured in the radio is $\sim$ 0.1\% for most Solar System comets \citep{Biv24}. Also, striking is the low CS abundance, which is in the low end of the range of values in Solar System comets.  On the other hand, CO/H$_2$O and CH$_3$OH/H$_2$O are enhanced with respect to the mean abundances in Solar System comets, while H$_2$CO/H$_2$O, and CH$_3$CN/H$_2$O are typical (Fig.~\ref{figabunH2O}).  From histograms of the abundance ratios with respect to HCN (Fig.~\ref{figabunHCN}), 3I/ATLAS displays high CO/HCN, H$_2$CO/HCN, CH$_3$OH/HCN, and CH$_3$CN/HCN abundance ratios and a rather low CS/HCN ratio. As concluded by \citet{2025arXiv251120845R}, 3I/ATLAS's coma is especially highly enriched in its CH$_3$OH/HCN abundance, being only surpassed by the peculiar CO-rich and N$_2$-rich comet, C/2016 R2 (PanSTARRS). It would be tempting to conclude to a high C/N ratio in 3I/ATLAS nuclear ices, but HCN is not the dominant carrier of nitrogen in cometary ices \citep{Biv24}. Figure~\ref{figS_surC} plots an histogram of the ratio of the total abundance of sulfur-bearing species to the total abundance of carbon bearing molecules in comets, considering only the CO, CH$_3$OH, H$_2$CO, CS and H$_2$S molecules observed in 3I/ATLAS. The value for 3I/ATLAS is at most the minimum value observed in comets, illustrating the peculiar chemistry of this interstellar object.

From the observation of the HDO line at 241.56 GHz (Table~\ref{tabarea}), we derived an upper limit for the D/H in H$_2$O of $<$ 0.91\% (at 3$\sigma$). This value is a factor of 10 or more higher than the value measured in comets \citep[e.g.][]{2022A&A...662A..69M,Biv24,2025NatAs...9.1476C}. For CH$_3$OH, we obtained an upper limit on the D/H value of 5.3\% by stacking 17 CH$_3$OD lines between 226 and 272 GHz. This is a factor of 4 larger than the value measured in comet 67P \citep{2021MNRAS.500.4901D}, thereby excluding extreme deuteration levels for this molecule in 3I/ATLAS. 

Perharps, the most surprising property of 3I/ATLAS's coma is the very low  expansion velocity, which is a factor of two below the value of $\sim$ 0.7--0.8 \kms~measured in comets of similar activity levels and at comparable heliocentric distances. This low velocity might be the signature of a gas flow dominated by heavy molecules such as CO$_2$ (Fig.~\ref{figvelo}). This molecule was found to be overabundant in 3I/ATLAS at 3.3 au \citep{2025ApJ...991L..43C}. 3I/ATLAS would then share similarities with the hyperactive comet 103P/Hartley 2, which nucleus released large amounts of CO$_2$ at its perihelion dragging water ice grains and chunks that subsequently sublimated in the coma \citep{2011Sci...332.1396A}. The amount of water and CO$_2$ molecules released from 103P/Hartley 2's nucleus were comparable \citep{2013Icar..225..688F}, which might explain why the gas expansion velocity in this comet \citep[0.6 \kms][]{2011Natur.478..218H} was not abnormally low (Appendix~\ref{appendix:velo}, Fig.~\ref{figvelo}). Alternatively, the low velocity measured in 3I/ATLAS could be direcly related to a large amount of coma gases released from icy grains. This is predicted by gas flow kinetic models including release both from the nucleus and from icy grains using a Direct Simulation Monte Carlo (DSMC) approach, and the effect is all the more significant that the amount of icy grains is high  \citep{2012Icar..221..174F}. 

 In conclusion, the observations of comet 3I/ATLAS with the IRAM 30-m telescope allowed to identify six molecules in its coma, with relative abundances which are at the extreme ends of the range of values measured in Solar System comets. Advanced modelling is needed to understand the unique dynamic properties of its atmosphere.


\begin{appendix}
\section{Log of the observations}

\begin{table*}
\caption[]{Log of observations.}\label{tablog}\vspace{-0.2cm}
\begin{center}
\begin{tabular}{rccccc}
\hline\hline
UT date  & $<r_{h}>$  & $<\Delta>$   & pwv & Integ. time & Freq. range$^a$ \\
$($yyyy/mm/dd.d--dd.d) & (au)  & (au) &    & (min)       & (GHz)  \\
\hline
2025/11/01.33--01.48 & 1.361 & 2.276 & 1.2~mm & 37 & 248.3-256.4, 264.3-272.4\\
        01.51--01.54 & 1.361 & 2.274 & 0.8~mm & 33 & 224.6-232.7, 240.6-248.7\\
        01.57--01.58 & 1.361 & 2.274 &  2~mm  & 16 & 146.8-154.9, 162.8-170.9\\
2025/11/02.34--02.45 & 1.364 & 2.265 & 12-5~mm & 94 & 146.8-154.9, 162.8-170.9\\
        02.47--02.51 & 1.364 & 2.264 &  3~mm  & 32 & 248.3-256.4, 264.3-272.4\\
        02.53--02.56 & 1.365 & 2.263 & 1.9~mm & 37 & 224.6-232.7, 240.6-248.7\\
2025/11/03.32--03.36 & 1.368 & 2.254 & 2.7~mm & 37 & 146.8-154.9, 162.8-170.9\\
        03.38--03.46 & 1.368 & 2.253 & 2.6~mm & 65 & 248.3-256.4, 264.3-272.4\\
        03.48--03.58 & 1.369 & 2.252 & 3-4~mm & 82 & 224.6-232.7, 240.6-248.7\\
\hline
\end{tabular}
\end{center}
\tablefoot{
\tablefoottext{a}{Frequency ranges covered in USB and LSB.}}
\end{table*}

\section{Detected lines and their characteristics}
\begin{table*}
\caption[]{Line intensities from IRAM observations}\label{tabarea} 
\begin{center}
\begin{tabular}{lllcclc}
\hline\hline
Date & Molecule & Transition  & Frequency       & Offset\tablefootmark{a} & $\int T_{mB} dv$ & Doppler shift \\
  $(yyyy/mm/dd.dd)$ &  &      & (MHz)        & (\arcsec) &       (K~\kms) & (\kms) \\
\hline
2025/11/02.36 & HCN      & $3-2$          & 265886.434   & 3.0 & $0.381\pm0.020$\tablefootmark{b} & $-0.04\pm0.03$  \\
2025/11/03.15 &          &              &              & 5.7 & $0.333\pm0.073$\tablefootmark{b} & $+0.17\pm0.13$  \\
2025/11/02.36 & HNC      & $3-2$          & 271981.142   & 3.0 & $0.047\pm0.018$ & $-0.32\pm0.26$ \\
2025/11/02.57 & CH$_3$CN & $8-7$$^c$ & 147149-147175 & 2.9 & $0.039\pm0.013$ &  \\ 
              &          & $9-8$$^c$ & 165540-165570 & 2.9 & $0.052\pm0.021$ &  \\
2025/11/02.57 & CH$_3$OH & $J_1-J_0 E$ ($J$=1--5) & 165050-165369 & 2.9 & $0.058-0.107(\pm0.009)$ &  $-0.07\pm0.01$  \\ 
2025/11/02.57 & CH$_3$OH & $J_1-J_0 E$ ($J$=6--9) & 165678-167931 & 2.9 & $0.078-0.041(\pm0.011)$ &  $-0.05\pm0.05$  \\ 
2025/11/02.57 & CH$_3$OH & $3_2-2_1 E$    & 170060.581  & 2.9 & $0.114\pm0.013$   & $-0.01\pm0.06$ \\ 
2025/11/02.40 & CH$_3$OH & $8_{-1}-7_0 E$ & 229758.811  & 2.9 & $0.109\pm0.012$  & $-0.01\pm0.06$ \\
2025/11/02.36 & CH$_3$OH & $5_K-4_K$$A^+, A^-, E^d$ & 241700-243916 & 2.9 & $0.101-0.337(\pm0.013)$ & $-0.03\pm0.01$ \\ 
2025/11/02.33 & CH$_3$OH & $J_3-J_2$$A^+$, $A^-$$^e$  & 251164-252253 & 2.9 & $0.027-0.068(\pm0.012)$ &  \\ 
2025/11/02.33 & CH$_3$OH & $11_0-10_1 A^+$ & 250507.016  & 2.9 & $0.068\pm0.012$   & $-0.04\pm0.09$ \\ 
2025/11/02.36 & CH$_3$OH & $5_2-4_1 E$    & 266838.123  & 2.9 & $0.199\pm0.012$  & $-0.04\pm0.03$ \\
2025/11/03.15 &          &              &             & 5.7  & $0.176\pm0.051$ & $+0.05\pm0.17$ \\  
2025/11/02.36 & CH$_3$OH & $9_0-8_1 E$    & 267403.394  & 2.9 & $0.056\pm0.016$  & $-0.20\pm0.17$ \\
2025/11/02.40 & CO       & $2-1$          & 230538.000   & 2.9 & $0.069\pm0.011$ & $-0.11\pm0.08$ \\
2025/11/02.36 & CS       & $5-4$        & 244935.557    & 2.9 & $0.040\pm0.012$   & $-0.36\pm0.20$   \\ 
2025/11/02.57 & CS       & $3-2$        & 146969.049    & 2.9 & $0.014\pm0.006$   & $+0.04\pm0.20$  \\ 
2025/11/02.40 & H$_2$CO  & $3_{12}-2_{11}$ & 225697.772  & 2.9 & $0.087\pm0.010$   & $+0.07\pm0.06$ \\ 
2025/11/02.57 & H$_2$CO  & $2_{11}-1_{10}$ & 150498.334  & 2.9 & $0.032\pm0.007$  & $+0.11\pm0.11$  \\ 
2025/11/02.36 & HCO$^+$     & $3-2$        & 267557.625  & 3.0 & $0.099\pm0.025$  & $+0.64\pm0.38$ \\
2025/11/02.57 & H$_2$S   & $1_{10}-1_{01}$ & 168762.762  & 2.9 & $0.015\pm0.011$   &  \\ 
2025/11/02.33 & SO       & $6,5-5,4$ & 251825.770  & 3.0 & $<0.036$   &  \\ 
2025/11/02.4  & (CH$_2$OH)$_2$ & $J=20-29$$^f$  & 225689-270647 &  3.0 &  $0.0039\pm0.0014$ & $-0.11\pm0.20$ \\
2025/11/02.4  & CH$_3$CHO   &  $J=12-14$$^g$  & 226552-271069 &  3.0 &  $0.0042\pm0.0011$ & $+0.19\pm0.13$ \\
2025/11/02.36 & HNCO     & $J=11-12, K=0,1$$^c$  & 240876-264694 &  2.9 & $0.051\pm0.023$ & $-0.11\pm0.26$ \\
2025/11/02.36 & HDO     &  $2_{11}-2_{12}$ & 241561.641 &  2.9 & $<0.033$ &  \\
\hline
\end{tabular}
\end{center}
\tablefoot{
\tablefoottext{a}{Pointing offset.} \tablefoottext{b}{Taking into account all hyperfine components.}
\tablefoottext{c}{Sum of four lines.} \tablefoottext{d}{7 lines.} \tablefoottext{e}{16 lines.}\tablefoottext{f}{45 lines.} \tablefoottext{g}{60 lines.} 
}\\
\end{table*}

\section{Synthetic line profiles}

\begin{figure*}
\begin{minipage}[c]{0.33\linewidth}
\resizebox{\hsize}{!}{\includegraphics[width=7cm,angle=270]{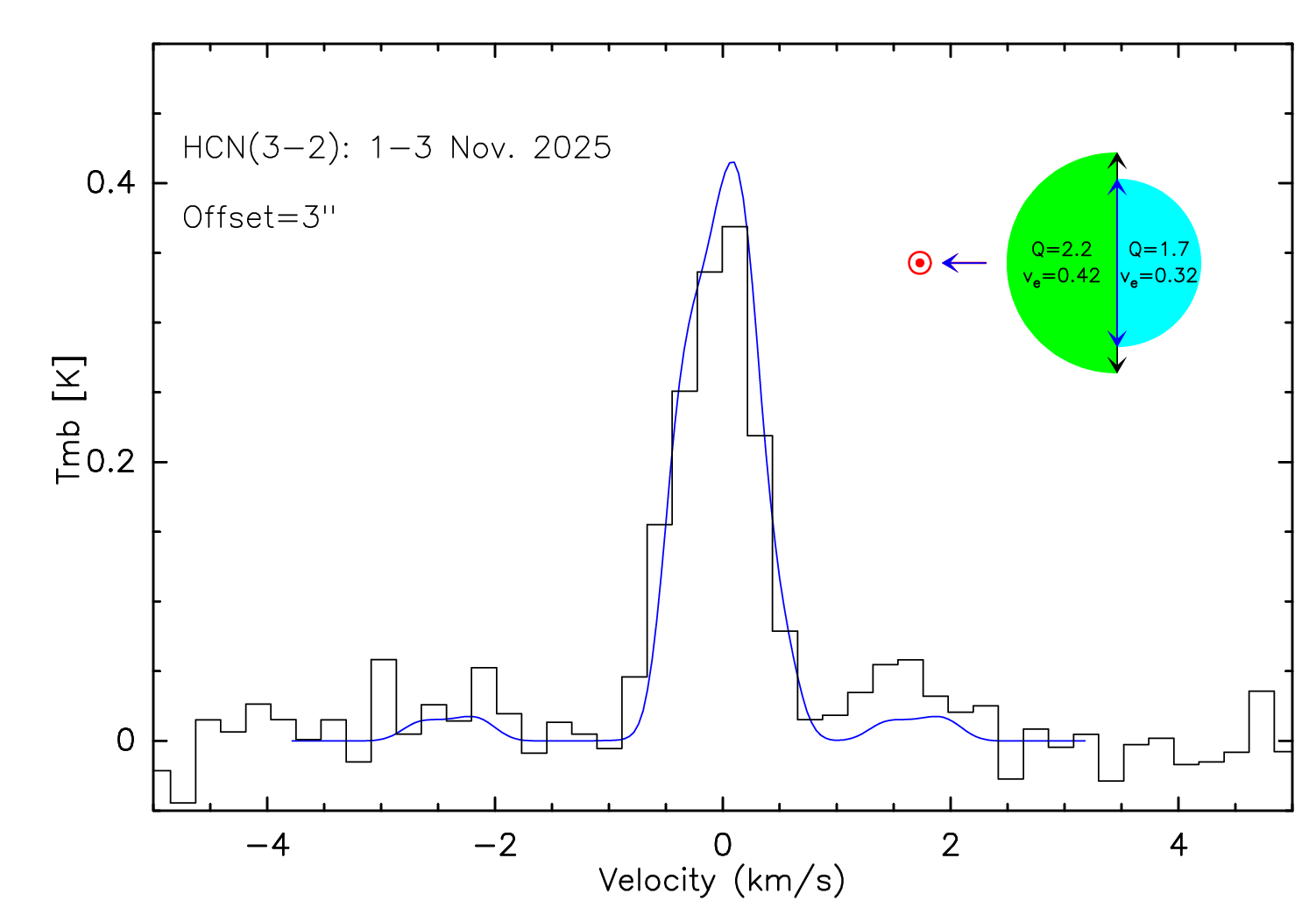}}
\end{minipage}\hfill
\begin{minipage}[c]{0.33\linewidth}
\resizebox{\hsize}{!}{\includegraphics[width=7cm,angle=270]{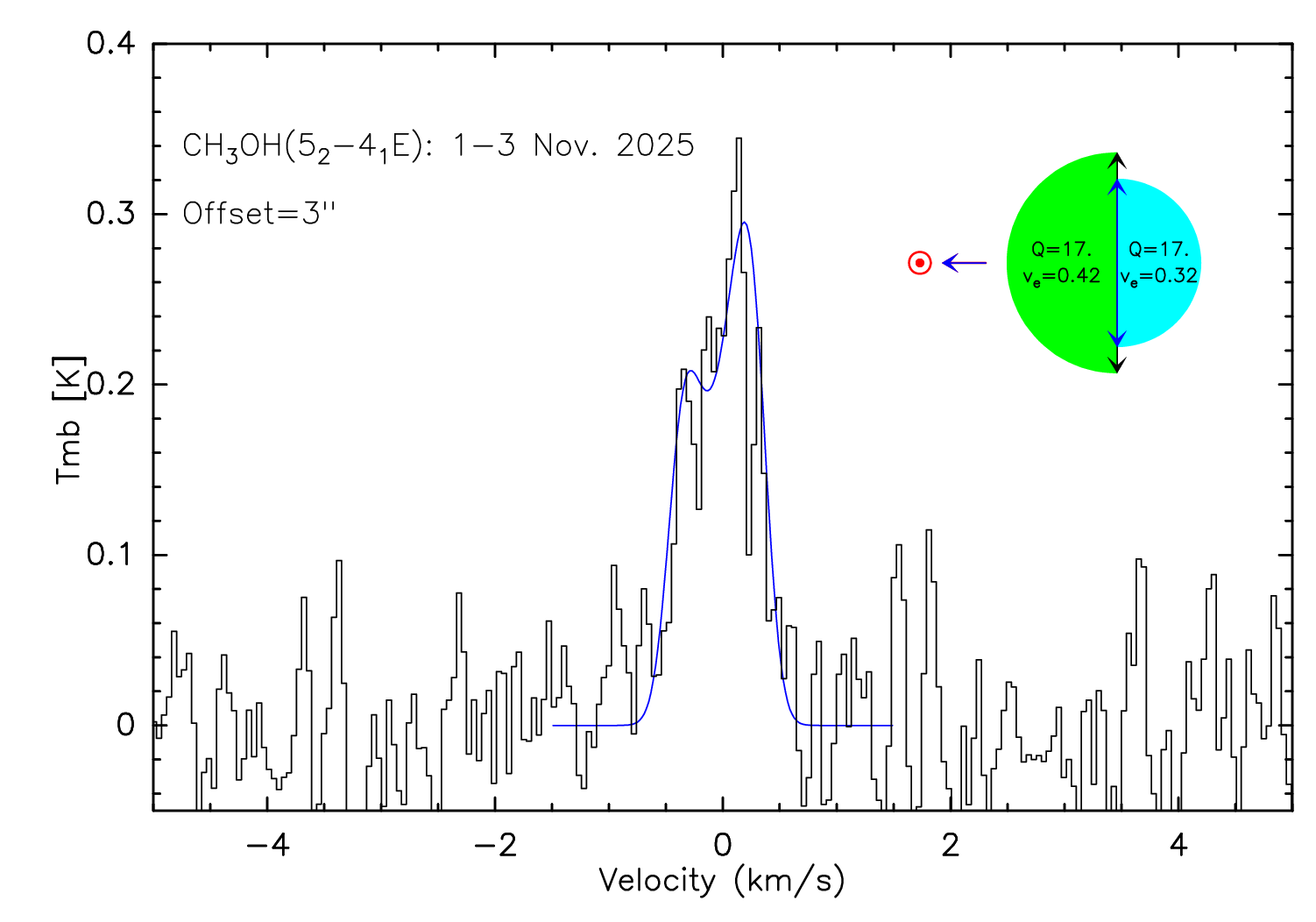}}
\end{minipage}\hfill
\begin{minipage}[c]{0.33\linewidth}
\resizebox{\hsize}{!}{\includegraphics[width=7cm,angle=270]{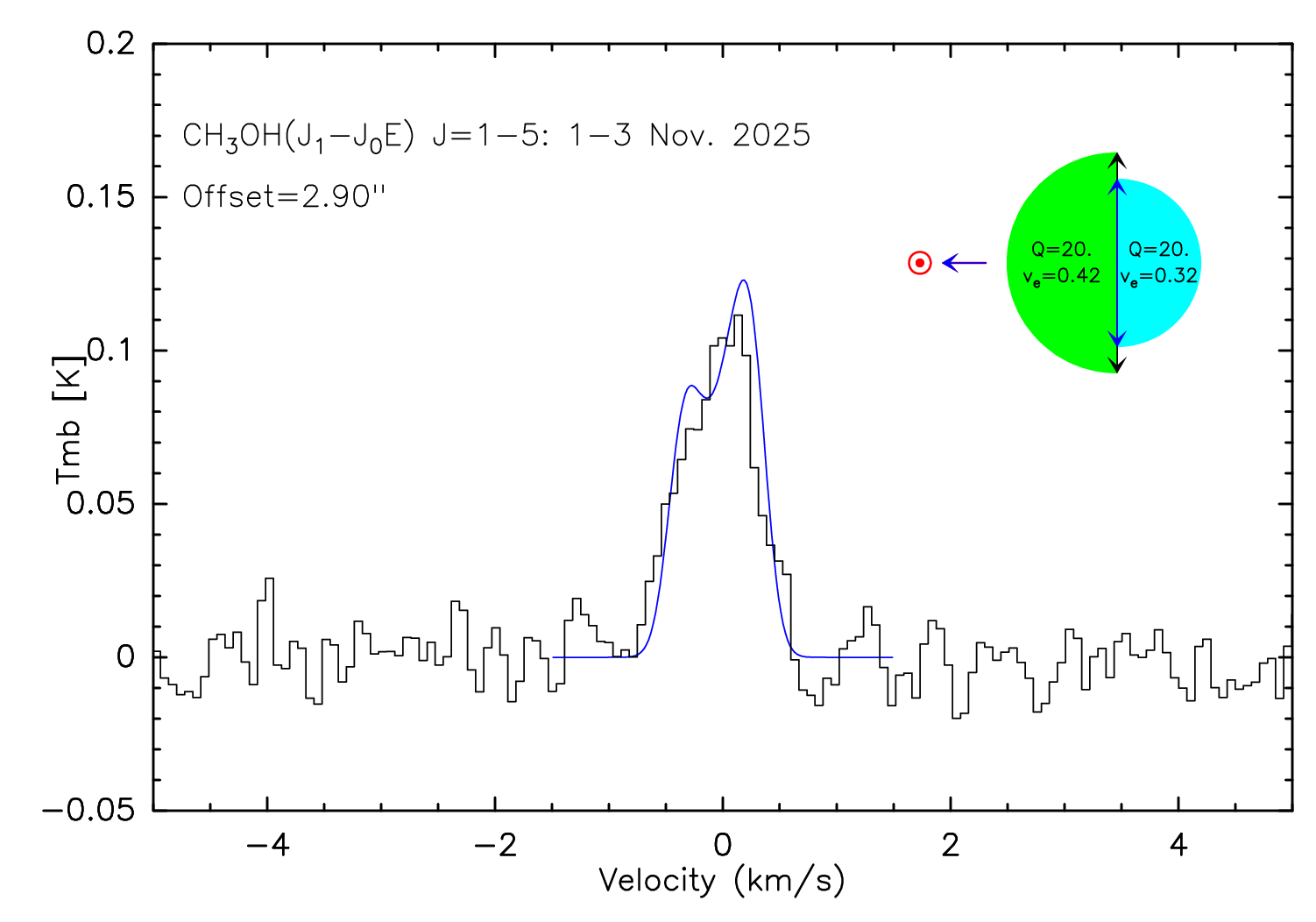}}
\end{minipage}
\caption{Synthetic line profiles (blue) superimposed on observed line profiles (black) of HCN (left panel) and CH$_3$OH (middle and right panels).
The inserts in the upper right of the plots show the assumed geometry of the outgassing with respect to the Sun direction, with the production rates given in units of 10$^{25}$ \mols~for HCN, and in units of 10$^{26}$ \mols~for CH$_3$OH. The assumed expansion velocities (\kms) in the sunward and antisunward hemispheres are also indicated. Thermal broadening assuming a kinetic temperature of 60 K (see main text) is considered. \label{linefits} } 
\end{figure*}

\section{CH$_3$OH rotation diagram}

\begin{figure*}
\sidecaption
\includegraphics[width=10cm,angle=270]{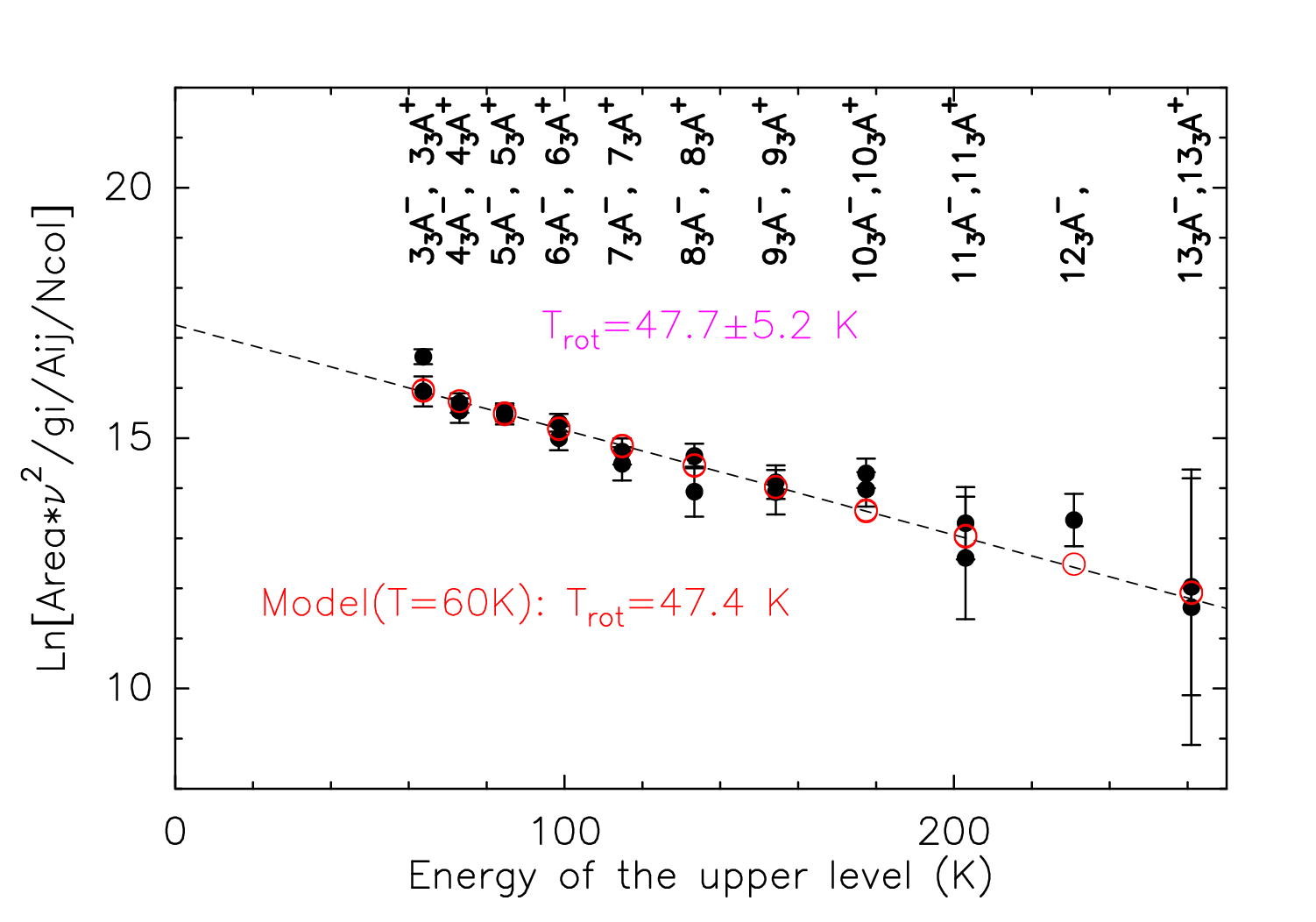}
\caption{Rotation diagram of CH$_3$OH, using $J_{\rm 3} -J_{\rm 3} A^+$ and  $A^-$ lines near 252 GHz. Red circles show the output of the excitation model assuming a kinetic temperature of 60 K. It reproduces nicely the measured rotational temperature of 47.7$\pm$5.2 K. The labels of the upper levels of the transitions are given at the top of the plot. \label{figdiag} } 
\end{figure*}

\section{Histograms of abundances relative to water}
\begin{figure}[h]
\resizebox{\hsize}{!}{\includegraphics[angle=270,width=12cm]{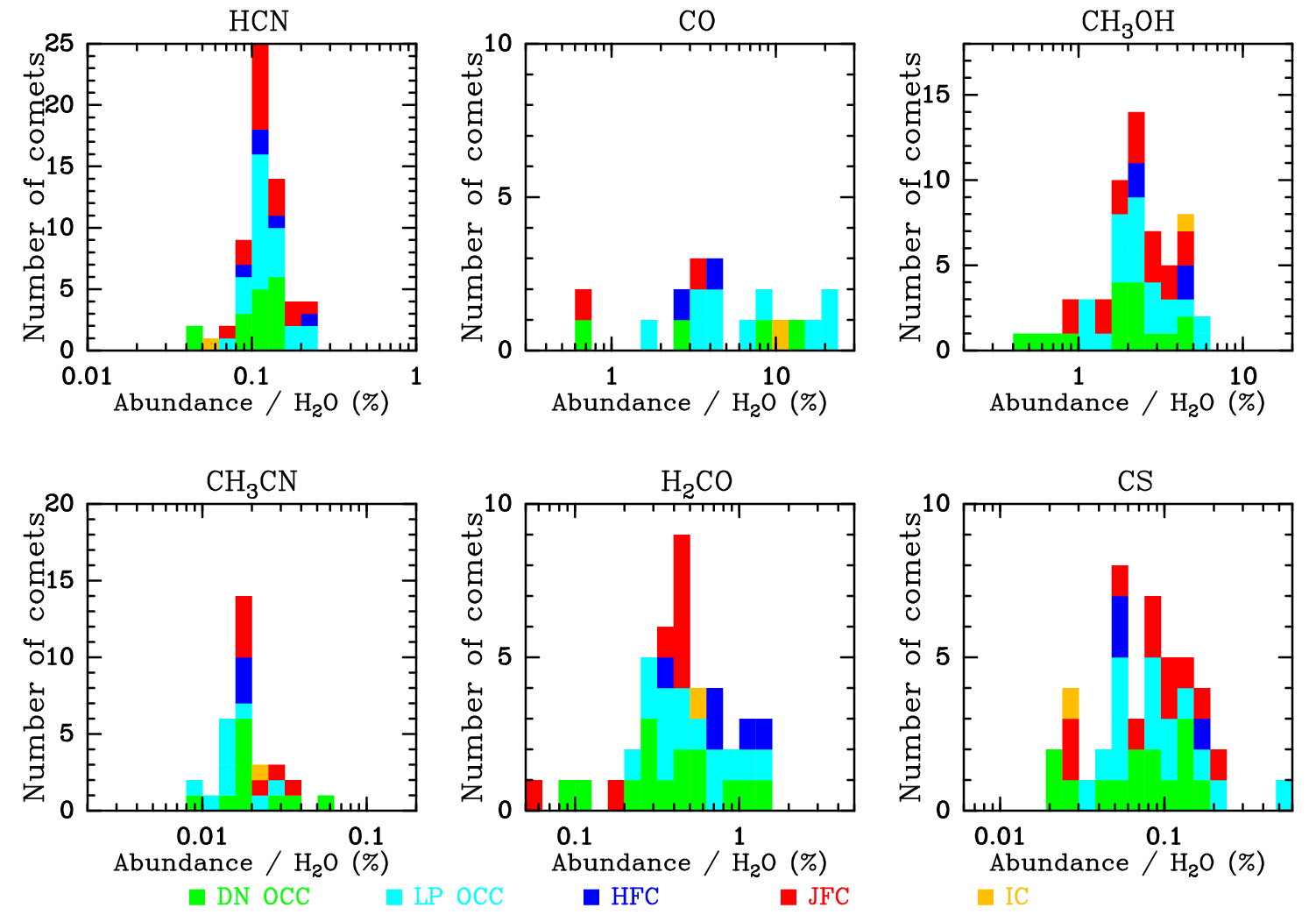}}
\caption{Same as Fig.~\ref{figabunHCN} for abundances 
relative to H$_2$O}
\label{figabunH2O}
\end{figure}

\section{Histogram of the Sulfur/Carbon ratio}
\begin{figure}[h]
\resizebox{\hsize}{!}{\includegraphics[width=12cm]{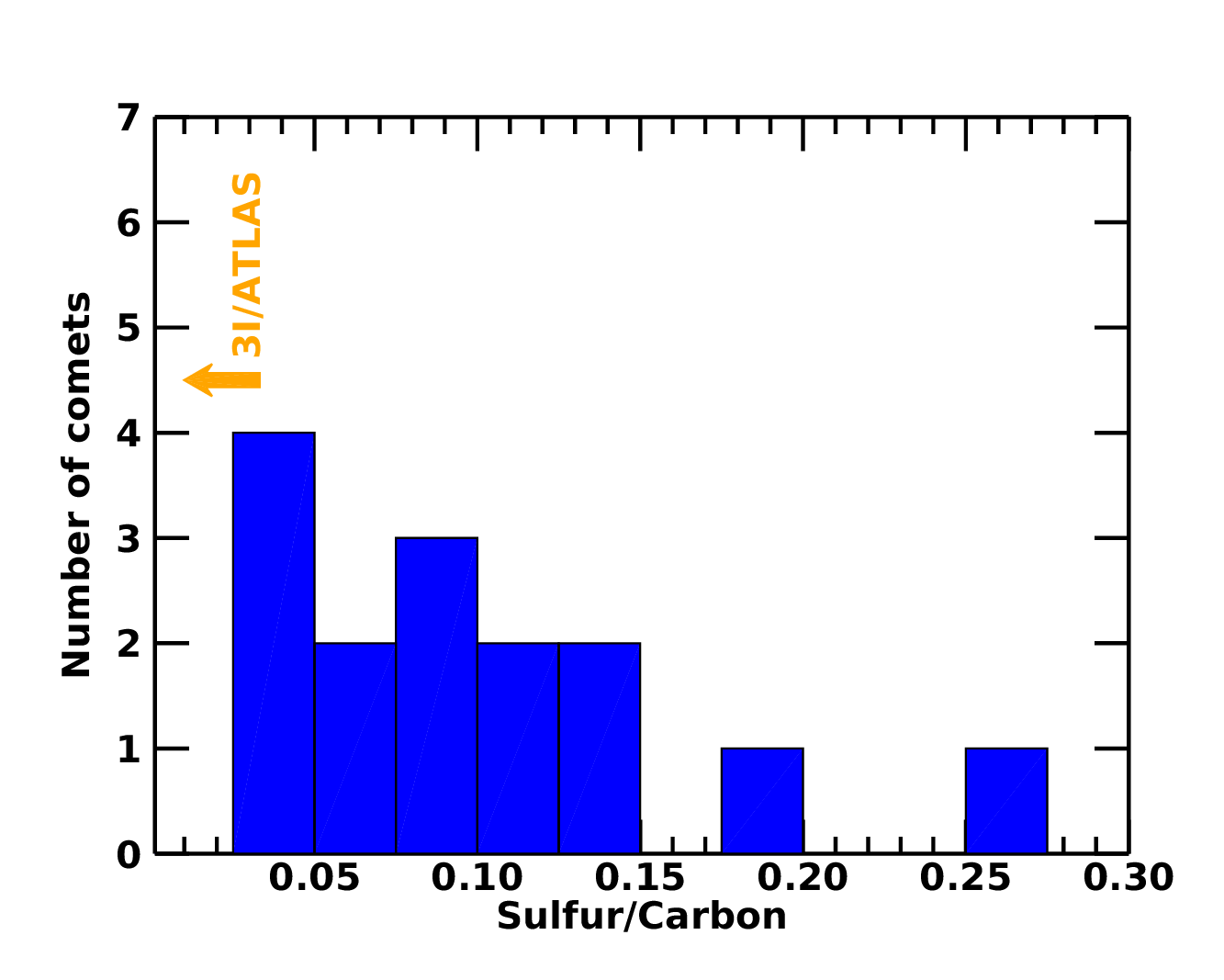}}
\caption{Histograms of the sulfur to carbon abundance ratio measured from observations at radio wavelengths. Considered sulfur and carbon-bearing molecules are CO, CH$_3$OH, H$_2$CO, H$_2$S and CS. Only comets for which these five species are detected are considered. Note however that H$_2$CO has a minor abundance, compared to CH$_3$OH and CO, and CS is much less abundant than H$_2$S in most comets. The upper limit for 3I/ATLAS is shown by an orange arrow. See \citet{Biv24} and references therein.  } 
\label{figS_surC}
\end{figure}

\section{Expansion velocity}
\label{appendix:velo}

\begin{figure*}
\sidecaption
\includegraphics[width=15cm]{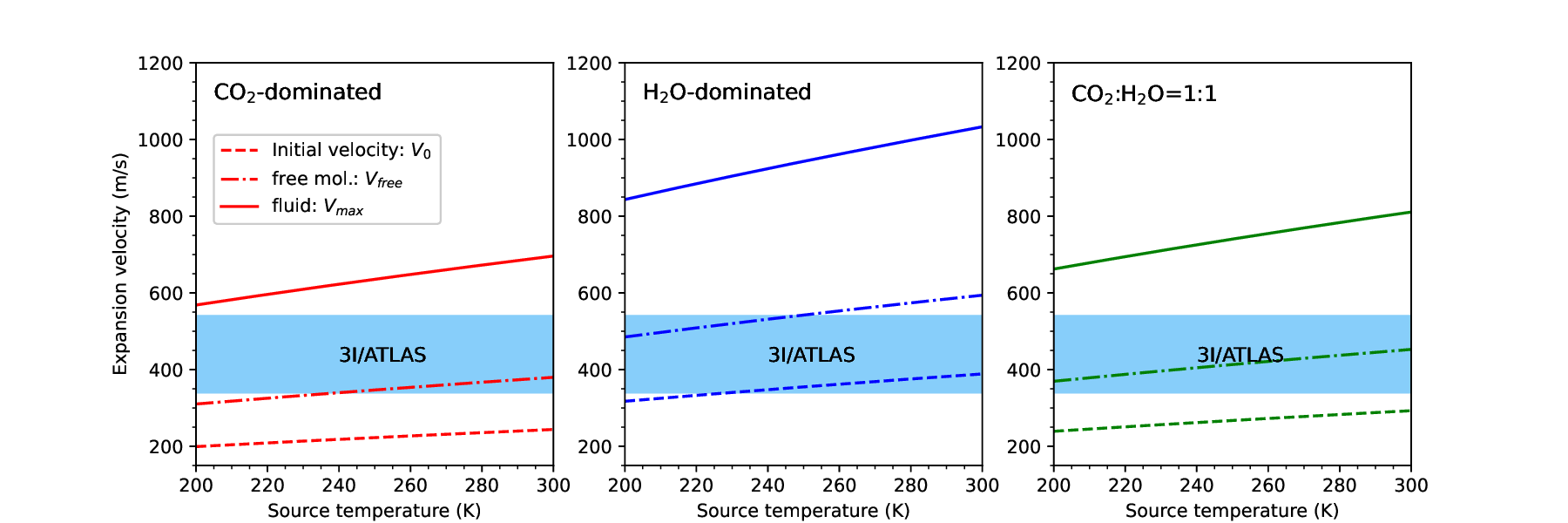}
\caption{Initial velocity $V_0$, maximal velocity $V_{\rm max}$, and terminal velocity for free-molecular expansion $V_{\rm free}$ as a function of source temperature ($T_N$). From left to right: CO$_2$-dominated, H$_2$O-dominated, and H$_2$O-CO$_2$ gas mixture. The range for 3I/ATLAS (in blue) delineates the expansion velocities measured in the blue and red wings of the line profiles (i.e. $VHM$ values, see main text).      } 
\label{figvelo}
\end{figure*}

Figure~\ref{figvelo} compares the expansion velocity measured in 3I/ATLAS from molecular line profiles to the gas velocity expected in three limiting cases.

We define the gas initial velocity $V_0$ as the sonic velocity:

\begin{equation}
V_0 = \sqrt{\frac{\gamma k_{\rm B} T_0}{m_{\rm g}}},
\end{equation}

\noindent
where $\gamma$ is the specific heat ratio (equal to 1.33 for H$_2$O, 1.28 for CO$_2$),  $T_0$ is the gas temperature at sonic line (equal to $\sim$ 0.82 times the temperature of the source releasing gases, referred to as $T_{\rm N}$), and $m_{\rm g}$ is the molecular mass of expanding gases. 

The maximal terminal gas velocity, which is reached in case of complete transfer of initial thermal energy into kinetic energy of the flow, is \citep[e.g.][]{Zakharov21}: 

\begin{equation}
V_{\rm max} = V_0 \sqrt{\frac{\gamma+1}{\gamma-1}}.
\end{equation}

\noindent
This terminal velocity is almost reached only for active comets, with large regions of fluid flow. This is shown by \citet{Zakharov23} who studied flow conditions from fluid to free molecular using a kinetic approach (Direct Simulation Monte Carlo (DSMC) method).

The maximum velocity reached in free-molecular expansion (i.e., in case of very weak gas production) is:

\begin{equation}
V_{\rm free} = \sqrt{\frac{8}{\pi} \frac{k_{\rm B} T_N}{m_{\rm g}}}.
\end{equation} 

In Figure ~\ref{figvelo}, we consider comas dominated by H$_2$O, by CO$_2$, and composed of both gases in equal quantities ($m_{\rm g}$ = 31 g/mole, $\gamma$ = 1.3).
Calculations are made for source temperatures from 200 K (temperature of exposed ice at typically $r_h$=1 au) to 300 K (appropriate if one considers gas diffusion through a layer of warm non-ice material). For an H$_2$O-dominated coma, $V_{\rm max}$ is close to the typical expansion velocity of 0.7--0.8 \kms measured in moderatly active comets ($Q_{\rm H2O}$ = 10$^{28}$--10$^{29}$ molec. s$^{-1}$). This outgassing regime is excluded for 3I/ATLAS.  The measured velocities are consistent with an H$_2$O-dominated outgassing only in the case of very weak water production (with free molecular expansion)(Fig.~\ref{figvelo}, central panel). A CO$_2$-rich coma is favored as in this case the expected gas expansion velocity is significantly lower (Fig.~\ref{figvelo}, left and right panels).

\begin{table}
\caption[]{Expansion velocities from line profiles.}\label{tableVHM}
\begin{center}
\begin{tabular}{lcc}
\hline\hline
Line & \multicolumn{2}{c}{Average half-widths\tablefootmark{a} ($VHM$)} \\
 & \multicolumn{2}{c}{(m s$^{-1}$)} \\
&  blue wing & red wing \\
\hline 
HCN 265.9 GHz & 463$\pm$33 & 387$\pm$33 \\
CH$_3$OH 266.8 GHz & 426$\pm$30 & 359$\pm$45 \\
CH$_3$OH 165 GHz & 672$\pm$59 & 305$\pm$35 \\
CO 230.5 GHz & 616$\pm$153 & 316$\pm$50 \\
\hline
\end{tabular}
\end{center}
\tablefoot{
\tablefoottext{a}{Values derived by fitting a half-Gaussian in the blue and red wings of the profiles. }
}\\
\end{table}

\end{appendix}


\begin{thebibliography}{}
\bibitem[A'Hearn et al.(2011)]{2011Sci...332.1396A} A'Hearn, M.~F., Belton, M.~J.~S., Delamere, W.~A., et al.\ 2011, Science, 332, 6036, 1396
\bibitem[Altwegg et al.(2019)]{Altwegg2019} Altwegg, K., Balsiger, H., \& Fuselier, S.~A.\ 2019, \araa, 57, 113
\bibitem[Bockel\'ee-Morvan et al.(2026)]{DBMMAJIS} Bockel\'ee-Morvan, D., Langevin, Y., Poulet, F.,  et al.\ 2026, The Astronomer's Telegram, 17726 
\bibitem[Biver(1997)]{1997PhDT........51B} Biver, N.\ 1997, Ph.D. Thesis, Universit\'e Paris VII 
\bibitem[Biver et al.(2021)]{2021A&A...648A..49B} Biver, N., Bockel{\'e}e-Morvan, D., Boissier, J., et al.\ 2021, \aap, 648, A49
\bibitem[Biver et al.(2024a)]{2024A&A...690A.271B} Biver, N., Bockel{\'e}e-Morvan, D., Handzlik, B., et al.\ 2024a, \aap, 690, A271
\bibitem[Biver et al.(2024b)]{Biv24} Biver, N. ; Dello Russo, N. ; Opitom, C. ; Rubin, M. 2024b, in Comets III, ed. K.J. Meech, M.R. Combi, D. Bockel\'ee-Morvan, S.N. Raymond \& M.E. Zolensky  (Space Science Series, University of Arizona Press), 459
\bibitem[Bodewits et al.(2020)]{2020NatAs...4..867B} Bodewits, D., Noonan, J.~W., Feldman, P.~D., et al.\ 2020, Nature Astronomy, 4, 867
\bibitem[Ceccarelli et al.(2023)]{2022arXiv220613270C} Ceccarelli, C., Codella, C., Balucani, N., et al.\ 2023, Protostars and Planets VII, 534, 379. 
\bibitem[Combi et al.(2026)]{Combi2025} Combi, M.~R., M{\"a}kinen, T., Bertaux, J.~L., et al.\ 2026, \apjl, 998, 1, L17
\bibitem[Cordiner et al.(2020)]{2020NatAs...4..861C} Cordiner, M.~A., Milam, S.~N., Biver, N., et al.\ 2020, Nature Astronomy, 4, 861
\bibitem[Cordiner et al.(2025a)]{2025NatAs...9.1476C} Cordiner, M.~A., Gibb, E.~L., Kisiel, Z., et al.\ 2025a, Nature Astronomy, 9, 1476
\bibitem[Cordiner et al.(2025b)]{2025ApJ...991L..43C} Cordiner, M.~A., Roth, N.~X., Kelley, M.~S.~P., et al.\ 2025b, \apjl, 991, 2, L43
\bibitem[Coulson et al.(2026)]{2025arXiv251002817C} Coulson, I.~M., Kuan, Y.-J., Charnley, S.~B., et al.\ 2026, \mnras, 546, 2, stag063
\bibitem[Crovisier et al.(2025)]{Cro25} Crovisier, J., Biver, N.,  Bockel\'ee-Morvan, D., 2025, {\it CBET} 5625
\bibitem[Deam et al.(2025)]{2025arXiv250705051D} Deam, S.~E., Bannister, M.~T., Opitom, C., et al.\ 2025, arXiv:2507.05051
\bibitem[Drozdovskaya et al.(2021)]{2021MNRAS.500.4901D} Drozdovskaya, M.~N., Schroeder I, I.~R.~H.~G., Rubin, M., et al.\ 2021, \mnras, 500, 4, 4901
\bibitem[Fougere et al.(2012)]{2012Icar..221..174F} Fougere, N., Combi, M.~R., Tenishev, V., et al.\ 2012, Icarus, 221, 1, 174
\bibitem[Fougere et al.(2013)]{2013Icar..225..688F} Fougere, N., Combi, M.~R., Rubin, M., et al.\ 2013, Icarus, 225, 1, 688
\bibitem[Hartogh et al.(2011)]{2011Natur.478..218H} Hartogh, P., Lis, D.~C., Bockel{\'e}e-Morvan, D., et al.\ 2011, Nature, 478, 7368, 218
\bibitem[Hui et al.(2026)]{2026arXiv260121569H} Hui, M.-T., Jewitt, D., Mutchler, M.~J., et al.\ 2026,  \apjl, 999, L37
\bibitem[Jehin et al.(2025)]{2025ATel17515....1J} Jehin, E., Hmiddouch, S., Aravind, K., et al.\ 2025, The Astronomer's Telegram, 17515, 1 
\bibitem[Kareta et al.(2020)]{2020ApJ...889L..38K} Kareta, T., Andrews, J., Noonan, J.~W., et al.\ 2020, \apjl, 889, 2, L38
\bibitem[Kuan et al.(2025)]{2025CBET.5628....1K} Kuan, Y.-J., Chuang, Y-L., Coulson, I.~M., et al.\ 2025, CBET, 5628 
\bibitem[Lin et al.(2020)]{2020ApJ...889L..30L} Lin, H.~W., Lee, C.-H., Gerdes, D.~W., et al.\ 2020, \apjl, 889, 2, L30
\bibitem[Lis et al.(2019)]{Lis2019} Lis, D.~C., Bockel{\'e}e-Morvan, D., G{\"u}sten, R., et al.\ 2019, \aap, 625, L5
\bibitem[Micheli et al.(2018)]{Micheli2018} Micheli, M., Farnocchia, D., Meech, K.~J., et al.\ 2018, Nature, 559, 223
\bibitem[M{\"u}ller et al.(2022)]{2022A&A...662A..69M} M{\"u}ller, D.~R., Altwegg, K., Berthelier, J.~J., et al.\ 2022, \aap, 662, A69
\bibitem[Opitom et al.(2019)]{2019A&A...631L...8O} Opitom, C., Fitzsimmons, A., Jehin, E., et al.\ 2019, \aap, 631, L8
\bibitem[Opitom et al.(2021)]{2021A&A...650L..19O} Opitom, C., Jehin, E., Hutsem{\'e}kers, D., et al.\ 2021, \aap, 650, L19
\bibitem['Oumuamua ISSI Team et al.(2019)]{2019NatAs...3..594O} 'Oumuamua ISSI Team, Bannister, M.~T., Bhandare, A., et al.\ 2019, Nature Astronomy, 3, 594
\bibitem[Roth et al.(2021)]{2021ApJ...921...14R} Roth, N.~X., Milam, S.~N., Cordiner, M.~A., et al.\ 2021, \apj, 921, 1, 14
\bibitem[Roth et al.(2026)]{2025arXiv251120845R} Roth, N.~X., Cordiner, M.~A., Bockel{\'e}e-Morvan, D., et al.\ 2026, \apjl, 999, L32
\bibitem[Seligman et al.(2025)]{2025ApJ...989L..36S} Seligman, D.~Z., Micheli, M., Farnocchia, D., et al.\ 2025, \apjl, 989, 2, L36
\bibitem[Tan et al.(2026)]{Tan2026} Tan, H., Yan, X., \& Li, J.-Y.\ 2026, \apjl, 998, 1, L22
 \bibitem[Zakharov et al.(2021)]{Zakharov21} Zakharov, V.~V., Rodionov, A.~V., Fulle, M., et al.\ 2021, Icarus, 354, 114091
 \bibitem[Zakharov et al.(2023)]{Zakharov23} Zakharov, V.~V., Rotundi, A., Bockel{\'e}e-Morvan, D., et al.\ 2023, Icarus, 395, 115453
\end{thebibliography}
\end{document}